\documentclass[11pt]{article}
\usepackage[margin=1.0in]{geometry}
\usepackage{graphicx}
\usepackage{epsfig}
\usepackage{amssymb,amsfonts,amsmath,amsthm}
\usepackage{mathrsfs}
\usepackage{epstopdf}
\usepackage{hyperref,color}
\usepackage{natbib}
\usepackage{setspace}
\usepackage[multiple]{footmisc}

\pdfoutput=1

\begin{document}

\title{The emergence of space and time}
\author{Christian W\"uthrich\thanks{I thank Robin Hendry and Tom Lancaster for their insightful and challenging comments on an earlier draft of this paper. This work was partly performed under a collaborative agreement between the University of Illinois at Chicago and the University of Geneva and made possible by grant number 56314 from the John Templeton Foundation and its content are solely the responsibility of the author and do not represent the official views of the John Templeton Foundation.}}
\date{6 April 2018}
\maketitle

\noindent
{\small For Sophie Gibb, Robin Finlay Hendry and Tom Lancaster (eds.), {\em Routledge Handbook of Emergence}.}


\begin{abstract}\noindent
Research in quantum gravity strongly suggests that our world in not fundamentally spatiotemporal, but that spacetime may only emerge in some sense from a non-spatiotemporal structure, as this paper illustrates in the case of causal set theory and loop quantum gravity. This would raise philosophical concerns regarding the empirical coherence and general adequacy of theories in quantum gravity. If it can be established, however, that spacetime emerges in the appropriate circumstances and how all its relevant aspects are explained in fundamental non-spatiotemporal terms, then the challenge is fully met. It is argued that a form of spacetime functionalism offers the most promising template for this project.
\end{abstract}

\noindent
Space and time, it seems, must be part and parcel of the ontology of any physical theory; of any theory with a credible claim to being a {\em physical} theory, that is. After all, physics is the science of the fundamental constitution of the material bodies, their motion in space and time, and indeed of space and time themselves. Usually implicit, Larry \citet{skl83} has given expression to this common intuition:
\begin{quote}
What could possibly constitute a more essential, a more ineliminable, component of our conceptual framework than that ordering of phenomena which places them in space and time? The spatiality and temporality of things is, we feel, the very condition of their existing at all and having other, less primordial, features... 
We could imagine a world without electric charge, without the atomic constitution of matter, perhaps without matter at all. But a world not in time? A world not spatial? Except to some Platonists, I suppose, such a world seems devoid of real being altogether. (45)
\end{quote}
The worry here, I take it, goes beyond a merely epistemic concern regarding the inconceivability of a non-spatiotemporal world; rather, it is that such a world would violate some basic necessary condition of physical existence. It is contended that space and time partially ground a material world. The alternative to a spatiotemporal world, it is suggested, is a realm of merely abstract entities.\footnote{The defenders of the claim that the world is purely abstract, formal, or mathematical---as opposed to {\em partly} abstract, formal, or mathematical---are usually referred to as `Pythagoreans', rather than as `Platonists'.} Part of what it means to be `physically salient' \citep{hugwut13b} is to be in space and time. In other words, what it is to give a physical explanation of aspects of our manifest world is, among other things, to offer a theory of how objects are and move in space and time. 

However, it turns out that physics itself may lead us to the conclusion that space and time are not part of the fundamental ontology of its best theories. Just as the familiar material objects of our manifest world arise from qualitatively rather different basic constituents, spacetime may emerge only from the collective action of fundamental non-spatiotemporal degrees of freedom. Spacetime, in other words, may exist merely `effectively', just as many salient aspects of our physical world, such as temperature, pressure, or liquidity. 

The present essay investigates this possibility. To this end, I precisify the situation as I see it (\S\ref{sec:emerge}) and illustrate it with concrete cases (\S\ref{sec:concrete}). In \S\ref{sec:function}, I articulate and defend the position of `spacetime functionalism' as the appropriate position to take vis-\`a-vis the suggestion that spacetime is emergent. Conclusions are in \S\ref{sec:outlook}.

\section{The emergence of space and time in fundamental physics}\label{sec:emerge}

There are two areas in physics in which space or spacetime may emerge from something qualitatively rather different: wave function monism in quantum mechanics and quantum theories of gravity. The former case concerns a particular interpretative issue in non-relativistic quantum mechanics. Although this in itself does not solve the measurement problem, part of such solution will be to get clear on the ontological status of the wave function. A central part of the formalism of quantum mechanics, the wave function mathematically expresses the state of a quantum system as the system's configuration. Governed by the Schr\"odinger equation, it is a function whose domain is a mathematical space whose dimensions are the chosen degrees of freedom (and time), normally the system's position in space. If the system at stake consists in $N$ particles moving in three-dimensional space, then the wave function's domain is the $3N$-dimensional configuration space. 

All interpretations of quantum mechanics are faced with the question of the ontological status of the wave function; and I will not be concerned with this general problem here. Suffice it to mention that it has been proposed \citep{alb96} that the wave function is the \emph{only} thing in the ontology of quantum mechanics. This position---`wave function monism'---faces the challenge of explaining the features of the manifest world, which seem very distant from the object described by the wave function. The issue, then, is to understand how three-dimensional physical space and what goes on in it emerges from the wave function in its high-dimensional configuration space. Thus, wave function monism must recover the spatiality of our physical world and its three-dimensionality. David \citet{alb15} defends a functionalist response on behalf of the wave function monist: ordinary objects in three-dimensional space are first `functionalized' in terms of their causal roles, i.e., reduced to a node in the causal network of the world; then it is argued that the wave function dynamically enacts these causal relations and, more generally, these functional roles precisely in a way which gives rise to the empirical evidence we have and to the manifest world more generally.\footnote{\citet{ney15} offers the most up-to-date evaluation of functionalism in the context of wave function monism.}

The second area concerns quantum theories of gravity.\footnote{See the collection by \citet{ori09} for a useful overview of approaches to quantum gravity.} Although general relativity stands unrefuted as our best theory of gravity and hence of spacetime, it assumes that matter exhibits no quantum effects. Since this assumption is false, it must ultimately be replaced by a more fundamental, i.e., more accurate and more encompassing, theory of gravity which takes the quantum nature of matter into full account. There are many proposals for how to articulate such a quantum theory of gravity, and unfortunately no empirical constraints to guide the search other than those confirming our currently best theories, such as the standard model of particle physics and general relativity. Saving concrete examples for \S\ref{sec:concrete}, an analysis of different research programs into quantum gravity reveals that in almost all of them, spacetime is absent from the fundamental ontology. More precisely, the ontology of these theories seems to consist of physical systems with less-than-fully spatiotemporal degrees of freedom. In other words, the structures postulated by these theories lack several, or most, of those features we would normally attribute to space and time, such as distance, duration, dimensionality, or relative location of objects in space and time. In short, spacetime, either as it figures in general relativity or phenomenologically in the experience of the world, is \emph{emergent} \citep{hugwut}.

In both cases, we witness the emergence of physical space or spacetime from something non-spatial or non-spatiotemporal.\footnote{Although of course the fundamental structures or their states may well be described by representations which inhabit a \emph{mathematical} space, such as a configuration space or a Hilbert space.} However, there are also significant disanalogies. For starters, space and time play very different roles in quantum mechanics: while position is an observable with a corresponding operator, time is the dynamical parameter entering the wave function and hence the Schr\"odinger equation. Departing from this difference between space and time, Alyssa \citet[\S7]{ney15} notes that the wave function monist has the dynamical evolution of the wave function available for their functionalist reconstruction project, which is generally not the case in quantum gravity, where time is part of the spacetime structure that evaporates at the fundamental level. Ney argues that this difference is responsible for the differential success of the two cases in dealing with the threat of empirical incoherence. A theory is \emph{empirically incoherent} just in case its truth undermines conditions necessary for its empirical confirmation. Ney maintains that since empirical confirmation is ineliminably diachronic but not spatial, fundamental time---and change---is essential for saving a theory from being empirically incoherent in a way that space is not. She concludes that, unlike wave function monists, quantum theories of gravity without time face imminent empirical incoherence. However, it is not clear on what basis such a distinction can be made; as long as space and time, which both seem necessary for empirical confirmation, can be shown to emerge at the appropriate scales of human science, the threat is averted in either case \citep{hugwut13b}. Be this as it may, it is clear that in quantum gravity it is \emph{spacetime}, and not mere \emph{space}, which is emergent rather than fundamental, and this constitutes the first important disanalogy.

There is a second crucial disanalogy. While it is mandatory for any interpretation of quantum mechanics to pronounce itself on the ontological status of the wave function, the content of that pronouncement need not be that the wave function gets reified and admitted to the ontology. Instead, one can opt for a `primitive ontology' of local `beables', i.e., basic entities in space and time \citep{alleal08}. This primitive ontology comprises the basis which makes up the entire world, including empirical evidence, and is postulated prior to an interpretation of the mathematical formalism. Only once the primitive ontology is set up, we check whether the formalism of the theory commits us to further entities. Among Bohmians, for instance, it is then popular to interpret the wave function as part of the \emph{nomological}, rather than the \emph{ontological}, structure of the world \citep{mil14,cal15a}. While wave function monism is thus clearly optional, it seems as if the fundamental absence of space and time from the ontologies of quantum gravity is not a matter of interpretation, and thus unavoidable. Furthermore, it is generic, i.e., present in most or all approaches to quantum gravity, as can be seen, among others, from the examples to be discussed in \S\ref{sec:concrete}. 

Turning to quantum gravity for the remainder of the essay then, what exactly do we mean when we say spacetime `emerges' from some non-spatiotemporal reality? Before anything else, it should be emphasized that quantum gravity is very much a work in progress, and hence we cannot hope to command a conclusive understanding of the situation just yet. In that sense, the term `emergence' serves as placeholder for a relation the investigation of which is part of the very project of quantum gravity: it is, as it were, a working title for that relation.

Nevertheless, there are a few things we can presently say, however provisional they may be. First, what are the relata of the `emergence' relation? What we are ultimately interested in are two aspects of the emergence of spacetime. First, the relationship between general relativity and the quantum theory of gravity must be understood. So it designates a relation between \emph{theories}. Second, it also refers to a relation that is supposed to generically hold between the physical structures of quantum gravity and relativistic spacetimes. Let us discuss these two aspects in turn. 

Qua relation between theories, emergence designates an asymmetric relation obtaining between general relativity and a quantum theory of gravity capturing several important aspects. First, it is a relation of \emph{relative fundamentality}, which partially orders scientific theories \citep[\S2]{wut12x}. A putative theory of quantum gravity is more fundamental than general relativity, although it may not be fundamental \emph{tout court} \citep{crolin18}. Second, it should be expected to be a \emph{broadly reductive} relation, both in the ontological and the explanatory sense. The ontology of general relativity should be expected to depend on, and hence be reducible to, that of a quantum theory of gravity. The scientific explanations at the level of general relativity should ultimately also be understood in terms native to the more fundamental theory, although this may be less obvious than for the ontology. This reducibility currently remains a promissory note. Although not strictly necessary, it would constitute a central part of an explanation in quantum gravity for the predictive successes (and failures, should there be) of general relativity. In this sense, reduction can be seen as a methodological directive in explicating the relation of emergence in the present case. This means that the form of `emergence' at stake cannot be the strong form, which excludes reducibility \citep{cha06}. Finally, although broadly reductive, the relation should also turn out to be somewhat \emph{corrective} in the sense that the more fundamental theory should deliver predictions which are measurably improved over those of general relativity. 

Qua relation between the entities or structures postulated by the theories at the two levels, emergence is an equally asymmetric relation. Unlike the case of relating theories, here it will generally be \emph{many-to-one}, relating a potentially very large number of fundamental structures to one and the same relativistic spacetime. This is to be expected as some minor differences at the fundamental level should not lead to any discernible difference at the emergent level: recombine the quantum-gravitational structure slightly, and the same relativistic spacetime will be its best approximation at the effective level. In other words, fundamental structures multiply realize the less fundamental structure, though the fundamental structures are not different in kind. A second marked---and important---difference to relating theories is the fact that in general, there will be some model of the quantum theory of gravity, i.e., some physically possible fundamental structures, which will not give rise to anything like spacetime. Just as elementary particles may fail to combine to give rise to a carbon molecule, or a liquid, or a life, fundamental structures in quantum gravity may inauspiciously assemble such that \emph{no effective spacetime} results. Thus, the fundamental theory should be expected to contain some models which do not resemble our actual world at all. 

Philosophers may complain that this use of `emergence' is unjustified in the context of quantum gravity, particularly in light of the expected reducibility. I will not quarrel about words here, but hasten to add that such protest misses the fact that `emergence' here is supposed to express precisely the dual aspects of \emph{dependence} and \emph{independence} traditionally associated with emergence. On the one hand, it captures the anticipated complete ontological and, to some degree explanatory, dependence of the level of general relativity on the more fundamental one of quantum gravity and its structures. On the other hand, it enounces the fact that the spacetime structures are qualitatively distinct from those found at the more fundamental level. This qualitative independence can be cashed out in different currency, with varying strength. At its strongest end, it asserts full autonomy of the levels and assumes, e.g., the form of non-supervenience, or at least of irreducibility. Even in its weakest form, however, it gives voice to the novelty involved in the contention that the fundamental structures of quantum gravity are relevantly non-spatiotemporal. That such novelty in itself need not entail some irreducibility but is commensurate with an underlying reductive relation is a widespread view in philosophy of physics, and has been given an explicit statement and defence by Jeremy \citet{but11a}. Motivated by effective field theory, Karen \citet[particularly Ch.\ 2]{cro16} similarly argues that of the many concepts of emergence on the market, those operative in quantum gravity should be considered compatible with reduction. In recent unpublished work \citep{cro18}, she convincingly contends that it is the \emph{reductive aspect}, which should be considered necessary for the relation between a quantum theory of gravity and general relativity, and emergence---if it obtains at all---does so merely non-essentially and in ways that fully depend on the details of the approach to quantum gravity taken. 

If it does obtain, the emergence of spacetime (particularly in so-called `canonical' quantum gravity) can come in grades or ``levels'' of severity, or so Daniele \citet{ori18} suggests. At the mild end of the spectrum, we find the apparent disappearance of (space and) time exemplified by moving to a quantum description of the gravitational field and thus of the chronogeometric aspects of spacetime. At the next step removed, the fundamental structure may consist in `pre-geometric' degrees of freedom, which are not merely in a quantum superposition, but whose eigenstates do not afford a chronogeometric interpretation at all. At the final level, these pre-geometric degrees of freedom may dynamically combine into different continuum phases, only some of which are chronogeometric or spatiotemporal. In this case, there are phase transitions from one phase to another, including for example a transition from a non-spatiotemporal or pre-geometric one to a spatiotemporal or geometric phase---a `geometrogenesis'.

\section{Concrete examples of spacetime emergence}\label{sec:concrete}

The severity and the characteristics of the disappearance of spacetime differ from one approach to quantum gravity to another. This section illustrates both of these dimensions of spacetime emergence. To this end, I will discuss causal set theory as well as loop quantum gravity and some of its extensions. It should be noted that the emergence, and hence non-fundamentality, of spacetime is a rather generic feature of quantum gravity.\footnote{This is the case even in string theory \citep{wit96,hug17} and condensed matter approaches to quantum gravity \citep{bai13}.} In this sense, causal set theory and loop quantum gravity are merely representative examples of this. 

\begin{figure}
\centering
\epsfig{figure=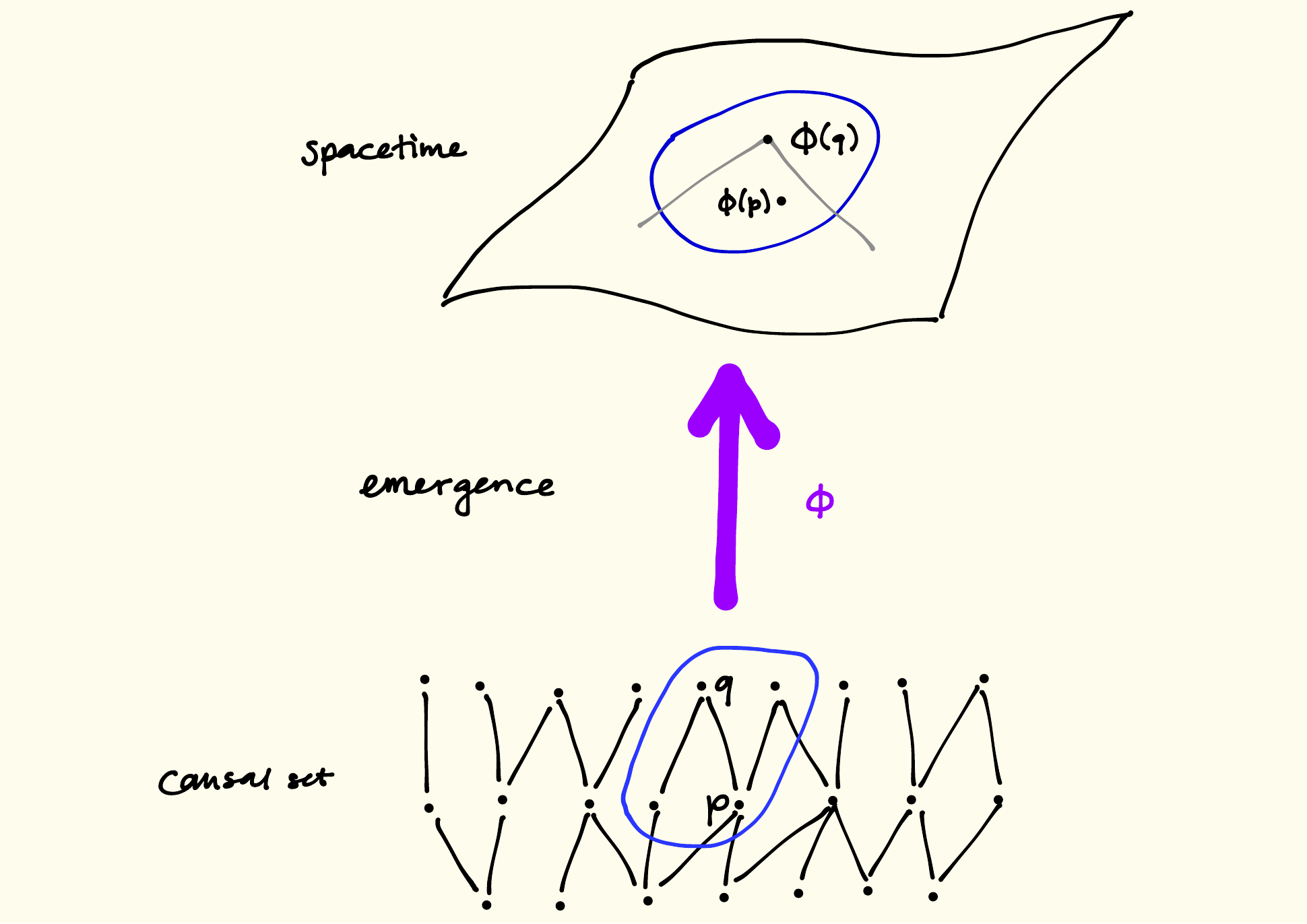,scale=0.65}
\caption{\label{fig:causet}A causal set and an emergent spacetime related by a map $\phi$.}
\end{figure}

Causal set theory starts out from a central result in general relativity according to which the causal structure determines, up to a so-called `conformal factor', the geometry of a relativistic spacetime. This result motivates causal set theory to postulate a discrete causal ordering as capturing the basic structure---a `causal set'. The causal ordering is a partial order formed by a fundamental relation of `causal precedence' on a set of otherwise unspecified elementary events, which is designed to be the discrete analogue of the causal (lightcone) structure of relativistic spacetimes. See Figure \ref{fig:causet} for an illustration of the resulting structure. 
As indicated in this figure, a relativistic spacetime may emerge from the fundamental causal set structure. A necessary condition for such emergence is that there exists a map $\phi$ from the elements of the causal set to spacetime events such that the causal structure is preserved: if an elementary event $p$ causally precedes another, $q$, then its image under $\phi$ must be in the causal past of $\phi(q)$, as in Figure \ref{fig:causet}. This condition, however, is far from sufficient, as generically causal sets give rise only to highly deformed spacetimes. The standard remedy is to impose additional conditions, whose task it is to rule out most of these unwieldy structures. The additional conditions are dynamical rules, which a causal set must follow as it grows from a finite past. Furthermore, since the causal set is a discrete structure and relativistic spacetimes are continua, emerging spacetimes cannot exhibit interesting features below the characteristic scale of the discrete structure of the causal set at stake. If it did, then the underlying causal set would not have the resources to ground them. Finally, a given causal set should, at least to a good approximation, give rise to no more than one emergent spacetime. This is imposed to ensure that while a single emergent entity can arise from distinct fundamental structures, one and the same fundamental structure cannot help but give rise to one specific emergent entity, if it does at all. The relation has to be many-to-\emph{one}, not many-to-many. Call this demand the `unique realization requirement'. 

The research program of causal set theory remains incomplete to date, in a number of respects. First, the unique realization requirement has not been shown to be satisfied in causal set theory. That it is satisfied is called the `Hauptvermutung', or main conjecture, by advocates of the research program. Its unproven status marks one of the program's main gaps. Second, the theory is so far entirely classical and as such cannot claim to meet the basic demand for a quantum theory of gravity. At the very least, it would have to offer a way of accommodating the quantum nature of matter. So far, it does not. 

Let us compare causal set theory and general relativity in terms of emergence as discussed in the previous section. It is clear that causal set theory is a candidate for a more fundamental theory than general relativity. Second, the ambition of the research program is to provide a reduction of general relativity, at least in the functional sense to be explicated in Section \ref{sec:function}. Should the reduction succeed, it will clearly be ontological and, again at least in the functional sense, explanatory. Third, although this can only be judged with the complete theory in place, the functional reduction of the relevant quantities is likely to result in small corrections for measurable spatiotemporal quantities, such as spatial distances and temporal durations. Furthermore, advocates of causal set theory claim that it predicts quantities, such as the cosmological constant, which remains a free parameter that has to be set by hand in general relativity.\footnote{\citet{sor91} takes the cosmological constant to result from statistical effects resulting from the large number of discrete bits. The `prediction' of the cosmological constant is severely limited, in several ways: first, it is off by at least an order of magnitude; second, it relies only on the scale of the fundamental discreteness, and not on any specifics of causal set theory; third, this scale is assumed to be the Planck scale, and thus also just put in `by hand'.} 

Turning to the relation between the structures rather than the theories, we do find that many distinct causal sets should give rise to one relativistic spacetime, and so the latter are multiply realizable by causal sets. Second, there are (too) many causal sets from which nothing spatiotemporal emerges: unless substantive dynamical constraints are added to the theory, most causal sets do not look like our world at all. 

Up to a few provisos, then, let us accept that many of the conditions for there to be the right kind of dependence of the general relativity on causal set theory are satisfied. But what about the \emph{independence}, or novelty, of the higher level from the lower level? Relativistic spacetimes (or general relativity) are independent from causal set (theory) insofar as the fundamental structures are not spatiotemporal, or at least not fundamentally so. As only the relation of causal connectibility is fundamental, a `spacelike' cross section of a causal set encompassed no relations among the basal elements whatsoever (cf.\ Fig.\ \ref{fig:causet}). Hence, what would be the best candidate to correspond to space is completely unstructured, and so \emph{a fortiori} does not have topological or metrical structure as we would expect a space to have. 

Although time is really only emergent qua aspect of the emerging spacetime structure, the causal relation ordering the fundamental structure bears clear similarities to time: it looks like a B-theoretic, discrete, relativistic version of an asymmetric temporal precedence relation. Once the full quantum character of the fundamental structure is implemented, however, one should expect to see these similarities erode and the independence become even more salient. Before I will comment on how the non-trivial task of recovering relativistic spacetimes can be accomplished, let me turn to loop quantum gravity, a case where we have at least a sketch of a quantum theory and where consequently the gap between the fundamental structures and spacetimes is wider. 

\begin{figure}
\centering
\epsfig{figure=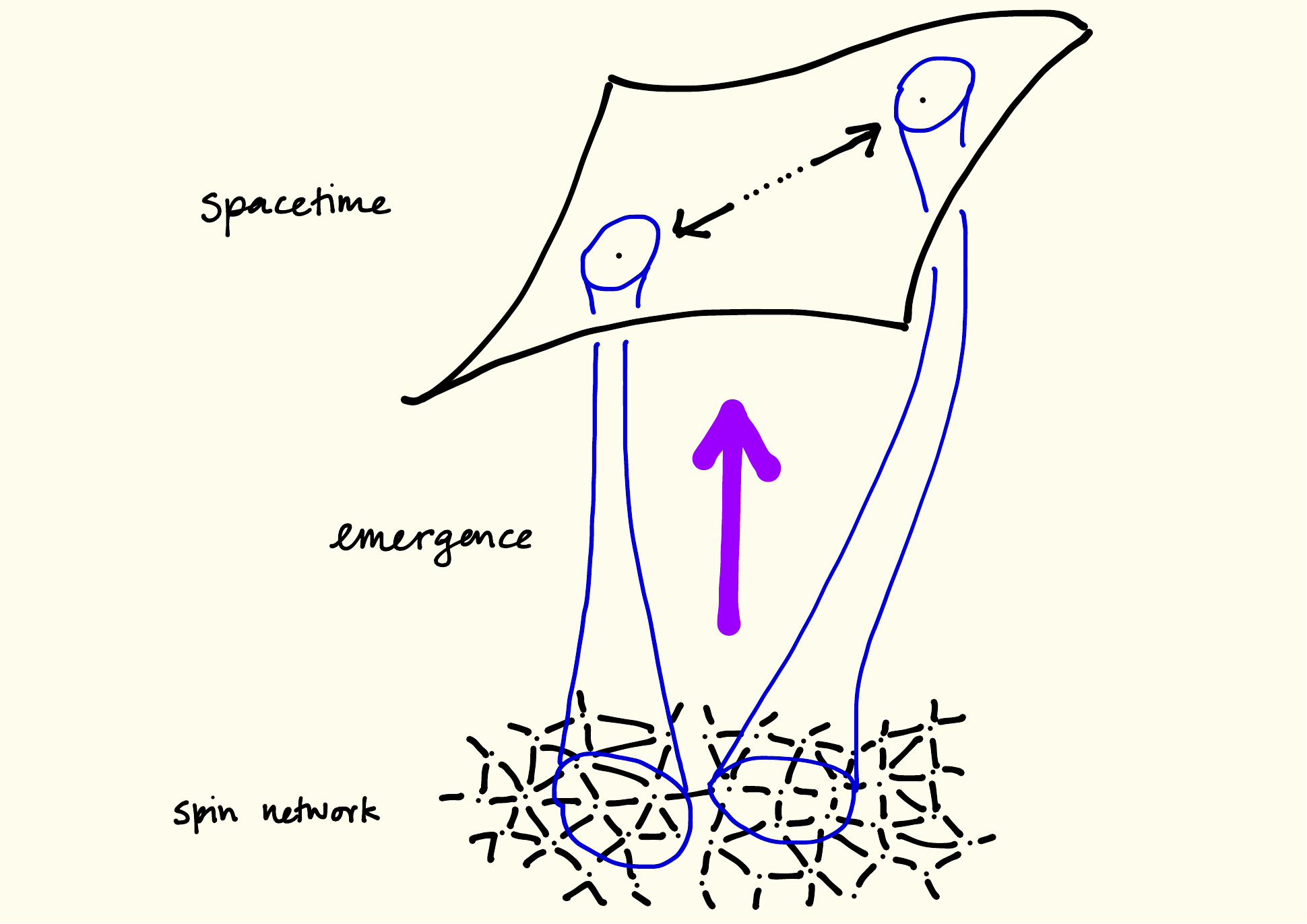,scale=0.6}
\caption{\label{fig:spinnetwork}A spacetime emerging from a spin network.}
\end{figure}

Loop quantum gravity also starts out from what it takes to be the central insights of general relativity and applies a standard quantization procedure in order to arrive at a quantum theory of gravity. Whatever the details of the quantization, the research program aims at the most conservative way to transform the dynamical geometry of general-relativistic spacetimes into quantum properties of a quantum system, relying on techniques that have proved successful on other occasions. The result, so far, is a quantum theory describing the fundamental structures that give rise to spacetime yet are so different from spacetime. These structures are spin networks, i.e.\ discrete structures consisting of nodes and connecting edges (Figure \ref{fig:spinnetwork}). Both the nodes and the edges have additional properties expressed by half-integer spin representations besides their connection in the network---hence `spin network'.

Before I address the question of time and dynamics, how should we interpret these spin networks? First, there is a sense in which they are `spatial': they result from a quantization of geometric structures that are standardly assumed to capture `space'. Second, the spin representations sitting on the nodes and the edges are eigenvalues of operators, which admit a natural geometric interpretation themselves: they represent three-dimensional volumes and two-dimensional surfaces, respectively. According to this interpretation, we thus obtain a compelling granular picture of spin networks: the spin representations  on the nodes depict the size of the volume associated with that `atom' of space, and the spin representations on the edges give the size of the area of the mutual surface of the two atoms connected by the edge. Thus, edges naturally express `adjacency' or contiguity between the `grains' of the discrete structure. 

However, the physical interpretation of spin networks as straightforwardly characterizing the elementary building blocks from which space is formed by joining the blocks together by the thread of adjacency is  limited in two ways. First, one finds a peculiar form of `non-locality' \citep{hugwut13b}: parts of the spin network that may be connected by an edge and thus are fundamentally adjacent may end up giving rise to parts of emergent spacetime which are spatially very distant from each other as judged by the distances operative at the level of emergent spacetime (the blue regions in Figure \ref{fig:spinnetwork} try to illustrate this non-locality). `Non-local' behaviour like this cannot be too rampant, for if it were, then the spin network in question would have given rise to a different spacetime, one which better reflects the fundamental locality structure. 

Second, it should be noted that the spin networks form a basis in the Hilbert space of the fundamental quantum systems at stake. If the state of such a system is in an exact spin network state, then the above geometric interpretation works well (modulo the `non-local' behaviour). However, in general the state of the system will be in a superposition of states in the spin network basis and thus not have any determinate geometric properties. Perhaps the state snaps into an eigenstate of the geometric operators upon measurement, but then the measurement problem of quantum theory rears its complicating head. Presumably, the measurement problem will be even less tractable in the present context, where we are dealing with a quantum system which is supposed to underwrite space(time)---all of it. 

What about the temporal aspect of spacetime? First, loop quantum gravity is a canonical approach to quantum gravity and as such faces the ominous `problem of time' \citep{ear02a,mau02,hugeal13}, better called `the problem of change'. This problem arises in canonical approaches: if we follow a standard way of interpreting physical theories, no genuinely physical magnitude can ever change---they are all `constants of the motion'. Following this line of thinking, then, the structures loop quantum gravity aims to quantize appear `frozen' in time. Furthermore, if one analyzes what ought to be the central dynamical equation (the so-called `Wheeler-DeWitt equation'), it appears as if no reference to time is made---rather odd for a dynamical equation expressing how a physical system is supposed to evolve over time! There exist various strategies to avoid or dissolve the problem, but the general conclusion seems to be that there is nothing naturally temporal in the fundamental structures described by loop quantum gravity.

In sum, then, the structures postulated by loop quantum gravity are non-spatiotemporal, arguably more so than causal sets. Relative to loop quantum gravity, general relativity enjoys a higher degree of novelty and thus independence than it did vis-\`a-vis causal set theory. Having said that, the fundamental structures described by some approaches to quantum gravity live on yet higher rungs of the ladder of increasing non-spatiotemporality as proposed by \citet{ori18}. For instance, group field theory \citep{ori14,hug18}, a `second quantization' of loop quantum gravity, proposes fundamentally algebraic, rather than primarily geometric, relations among the basic constituents. The structures can be in different phases, just like gases and condensates, and only when they `transition' into a spatiotemporal phase do they form structures such as spin networks that can give rise to the geometry of spacetime. Such a phase transition can obviously not be a dynamic process in time---it is a process `prior' to there being time as it were. Instead, it is captured by a renormalization group flow \citep{ori18cup}. 

Turning to the dependence aspects, loop quantum gravity is more fundamental and more encompassing than general relativity, with a relation between the theories that is broadly reductive. It potentially also offers deeper explanations and corrections of general relativity. For instance, although the expectation values of sufficiently large regions of space tend very strongly to the classical values, for sufficiently small regions, there ought to be noticeable differences between the values as predicted by loop quantum gravity and by general relativity. Furthermore, the measurement outcomes can only assume discrete values in loop quantum gravity. Perhaps closer to practice, loop quantum gravity promises to deliver explanations of phenomena connected with black holes or the very early universe, which remained unexplained in general relativity. As for the relation between the structures postulated by the theories at the two levels, we find again an instance of multiple realizability, as well as fundamental structures which fail to give rise to spacetimes.

\section{Spacetime functionalism}\label{sec:function}

To date, no research programme in quantum gravity has managed to fully recover general relativity and its spacetimes. Not only does thus remain a central explanatory task of a new scientific theory unfulfilled, but the threat of empirical incoherence as articulated above also looms over quantum gravity. The details are too rich and too technical to review here,\footnote{For those details, see \citet{wut17a,hugwut}.} but let me finish by sketching in broad philosophical outline what the task involves. 

First and foremost, this will involve the scientific work establishing a rather general mutual dependency under variations of the physics at the fundamental level and the spacetime physics at the phenomenal level. Will this suffice to discharge the task and to dispel the threat of empirical incoherence? According to the \emph{spacetime functionalist}, it does. Recently risen to popularity,\footnote{\citet{kno13} offers an statement and defence of spacetime functionalism in the context of classical gravity in terms of inertial frames, although it only receives this label in \citet{kno14}. \citet{lamwut18} apply the idea to the context of non-spatiotemporality in quantum gravity. Functionalism also offers an important perspective for wave function monism; cf.\ e.g.\ \citet[Ch.\ 6]{alb15}, \citet{ney15}, \citet[\S2]{lamwut18}.} functionalism asserts that in order to secure the emergence of spacetime from non-spatiotemporal structures in quantum gravity, it suffices to recover only those features of relativistic (or indeed phenomenal) spacetimes, which are functionally relevant in producing empirical evidence. Specifically, it is not necessary to recover the full manifold structure and the full metric of general relativity. If this is right, the threat of empirical incoherence is \emph{ipso facto} averted. 

In order to execute the functionalist program in quantum gravity, two steps would thus be necessary and sufficient: first, the spatiotemporal properties or entities are `functionalized' or `functionally reduced', i.e., they are given a definition in terms of their functional roles; second, an explanation is provided as to how the fundamental entities or structures found in quantum gravity successfully instantiate these functional roles. The functional roles that need to be filled include, among others, spacetime (relative) localization, spatial distances, temporal durations; the fundamental structures which enact these roles are parts of causal sets, or parts of spin network states, as the case may be. \citet[\S\S4-5]{lamwut18} sketch in more detail how this functionalist endeavour can be implemented in causal set theory and in loop quantum gravity. 

It should be clear that multiple fundamental structures can then in principle realize the relevant spacetime roles. This very liberal functionalist attitude raises concerns about the metaphysical robustness of the proposal. For instance, in analogy to the philosophy of mind, one may worry that the spacetime functionalist cannot muster the resources to account for the `spacetime qualia', i.e., the qualitative aspects of spacetime---the `spatiality' and the `temporality' of our experienced world---which seem phenomenologically vital and hence ontologically ineliminable. To make the objection salient, imagine the case of `spacetime zombies'. Suppose there are two worlds exactly alike in all non-spatiotemporal respects, with one of the worlds containing, additionally, exemplifications of spatiotemporal properties, which the other one lacks. Although their fundamental physics and all the functionally realizable spatiotemporal appearances are precisely the same, this latter world lacks genuinely spatiotemporal properties and instead only contains `spacetime zombies'. If this world is indeed numerically distinct from the first and metaphysically possible, it seems as if a merely functionalist treatment of spacetime does not command the resources to distinguish between the two. Hence, spacetime functionalism is false.

A more persuasive form of the worry has been raised by \citet{ney15}. In the context of quantum gravity, even though spacetime and spatiotemporal aspects of its occupants may be functionally reduced to fundamental, non-spatiotemporal physics, the objection goes, such functional reduction is insufficient in demonstrating how the fundamental structure do, in fact, \emph{constitute} spacetime and its spatiotemporal occupants. Similarly, David \citet{yat18} argues that functionalism is insufficient to solve the challenge of empirical incoherence, as our concepts for spatiotemporal properties are non-transparent in a fundamentally non-spatiotemporal world and functionalism cannot overcome this non-transparency. 

Against these detractors, \citet{lamwut18} argue that there remains no task left undone once a research programme can show ``how the fundamental degrees of freedom can collectively behave such that they appear spatiotemporal at macroscopic scales \emph{in all relevant and empirically testable} ways'' (10, emphasis in original). If this is right, then however counterintuitive a fundamentally non-spatiotemporal world may appear, it arises as a live possibility---one perhaps borne out in contemporary fundamental physics. This possibility also suggests that the widespread and intuitive `constitutionalism', i.e., the view according to which our experienced world is ultimately constituted by, or built up from, elementary building block or atoms of spatiotemporal `beables', may not offer an adequate understanding of the truly puzzling situations we presently face in fundamental physics. Instead, a functionalism devoid of constitutionalism and hence lighter on metaphysical prejudice is recommended.

\section{Conclusion and outlook}\label{sec:outlook}

Despite robust appearances to the contrary, research in quantum gravity suggests that the world we inhabit may not be fundamentally spatiotemporally. Philosophers should pay attention to these developments, as they have potentially profound implications for the philosophy of space and time, and beyond. If borne out, the absence of spacetime at the fundamental level forces at least a reconsideration of the traditional debate between substantivalism, relationalism, and structuralism. If substantivalism is reduced to the assertion that spacetime is an independent substance, and as such in no way derivative, then a substantivalist position could hardly be maintained. But relationalism does not obviously fare any better: although spacetime indeed seems to be derivative, it is not clear that it is derivative on material entities, let alone on the spatiotemporal relations in which these entities stand. The fundamental structures of quantum gravity are not obviously material, and they fail to exemplify spatiotemporal relations as directly as standard relationalism presupposes. The traditional position which is most naturally adapted to the present context is structuralism \citep{wut12c}, although of course the relevant structure in our ontology will not be straightforwardly spatiotemporal as is again assumed in the traditional formulation of spacetime structuralism.

Although the non-spatiotemporality varies in degree and character among the different research programmes, as was illustrated in \S\ref{sec:concrete}, the absence of spacetime from the ontology of quantum theories of gravity raises worries about the empirical coherence of these theories and, more generally, about their adequacy as replacements of general relativity. In order to resolve these concerns, however, it suffices to establish how spacetime emerges in the appropriate way from the fundamental non-spatiotemporal structures. This task consists in both technical and philosophical challenges that must be met in order to arrive at a satisfactory understanding. In \S\ref{sec:function}, I have suggested that a form of spacetime functionalism may furnish a key element to this project. But much work---including philosophical---remains to be done!

\bibliographystyle{plainnat}
\bibliography{}

\end{document}